\documentclass[preprint,prc,superscriptaddress,unsortedaddress,showpacs,preprintnumbers,amsmath,amssymb]{revtex4}
\usepackage{tipa}

\usepackage{graphicx}
\usepackage{dcolumn}
\usepackage{bm}

\def\beq{\begin{equation}}
\def\eeq{\end{equation}}
\def\bea{\begin{eqnarray}}
\def\eea{\end{eqnarray}}

\def\fun#1#2{\lower3.6pt\vbox{\baselineskip0pt\lineskip.9pt
  \ialign{$\mathsurround=0pt#1\hfil##\hfil$\crcr#2\crcr\sim\crcr}}}

\preprint{}

\begin{document}

\title{Angle-dependent Gap state in Asymmetric Nuclear Matter}

\author{Xin-le Shang}\email[ ]{shangxinle@impcas.ac.cn}
 \affiliation{Institute
of Modern Physics, Chinese Academy of Sciences, Lanzhou 730000,
China}
\author{Wei Zuo}
\affiliation{Institute of Modern Physics, Chinese Academy of
Sciences, Lanzhou 730000, China}\affiliation{State Key Laboratory of Theoretical Physics, Institute of Theoretical Physics, Chinese Academy of
Sciences, Beijing 100190, China}

\begin{abstract}
We propose an axi-symmetric angle-dependent gap (ADG) state with
the broken rotational symmetry in isospin-asymmetric nuclear
matter. In this state, the deformed Fermi spheres of neutron and
proton increase the pairing probabilities along the axis of
symmetry breaking near the average Fermi surface. We find the
state possesses lower free energy and larger gap value than the
angle-averaged gap state at large isospin asymmetries. These
properties are mainly caused by the coupling of different $m_{j}$
components of the pairing gap. Furthermore, we find the transition
from the ADG state to normal state is of second order and the ADG
state vanishes at the critical isospin asymmetry $\alpha_{c}$
where the angle-averaged gap vanishes.

\end{abstract}
\pacs{21.65.Cd, 26.60.-c, 74.20.Fg, 74.25.-q}

\maketitle

\section{Introduction}
The neutron-proton (n-p) pairing properties play an important role
in the description of superfluidity of finite nuclei with $N\simeq
Z$ \cite{Fn,Fn2} and symmetric nuclear
matter\cite{Snm1,Snm2,Snm3}. In general, the n-p pair correlations
are considered in different dominant partial-wave channels,
depending on the relevant density and temperature. For weakly
isospin-asymmetric systems, the isospin singlet attractive
$^{3}S_{1}-^{3}D_{1}$ ($^{3}SD_{1}$) channel dominates the pairing
interaction at relatively low densities around the nuclear
saturation density due to the tensor component of the nuclear
force\cite{Snm1,Sdd,Sdd2,Sdd3,Sdd4,Sdd5}, and the $^{3}D_{_{2}}$
channel dominates at high densities well above the saturation
density\cite{D1,D2}. In neutron star matter, the n-p pair
correlations are strongly suppressed by the isospin-asymmetry.
However, the dilute nuclear matter at sub-saturation densities in
supernovas and hot proto-neutron stars can support $^{3}SD_{1}$
channel pairing\cite{Sdd3,Sh1,Sh2,Sh3}.

Since n-p pair correlations depend crucially on the overlap
between the neutron and proton Fermi surfaces, the pairing gap
is suppressed rapidly as the system is
driven out of the isospin-symmetric state. At zero temperature, a small isospin-asymmetry is enough to prevent the formation of the
Cooper pairs between neutrons and protons with momenta $\overrightarrow{k}$ and $-\overrightarrow{k}$ around their average Fermi
surface where the contribution to superfluidity is dominant.
Near zero temperature, thermal excitations can reduce the
suppression by smearing out the two Fermi surfaces, however, it is
ineffective when the separation between the two Fermi surfaces is
large compared to the temperature. In isospin-asymmetric nuclear
matter, the FFLO \cite{ff,lo} state and the DFS (deformed Fermi
surfaces) \cite{dfs} state have been studied in
Refs.\cite{ffn,dfsn}. 
In a FFLO state,
the shift of the two Fermi spheres with respect to each other,
resulting form the collective motion of the Cooper pairs with a finite momentum,
enhances the overlap between the neutron and proton Fermi surfaces.
The overlap regions then provide the kinematical phase space for n-p
pairing phenomena to occur. And in a DFS state, the deformation
of the neutron and proton Fermi surfaces may increase the phase-space overlap between
the two Fermi surfaces. Both in these two kinds of
possible superfluid states, the quasiparticle excitation spectra
are no longer isotropic, since the anisotropic overlapping
configurations could increase the pairing energy. On the other
hand, the usually adopted angle-averaging procedure in the
previous calculations\cite{Sdd2,ffn}, which has been proved to be
a quite good approximation in symmetry nuclear matter \cite{aap},
considers the gap as an isotopic gap by ignoring the angle
dependence. As the true ground state corresponds to the anisotropic
overlapping configuration, the angle-averaging procedure may be an
insufficient approximation in isospin-asymmetric nuclear matter.

In this paper we consider an axi-symmetric angle dependent gap (ADG) state, and give a general and systematic comparison between the ADG state and the angle-averaged gap (AAG) state in isospin-asymmetric nuclear matter. The paper is organized as follows: In Sec. II we briefly review the formalism for the isotropic AAG state, and derive the angle
dependent gap equations from the Gorkov equations. The numerical solutions of these equations are shown and discussed in Sec. III, where we compare the AAG state with the ADG state at finite temperature. Finally, a summary and a conclusion are given in Sec. IV.



\section{Formalism}
For isospin-asymmetric nuclear matter, the isospin singlet $^{3}SD_{1}$ pairing channel dominates the attractive
pairing force at low densities. In this case we can consider $^{3}SD_{1}$ channel only, the gap function is thus expanded according to
\begin{eqnarray}
\Delta_{\sigma_{1},\sigma_{2}}(\textbf{k})
=\sum_{l,m_{j}}\Delta_{l}^{m_{j}}(k)[G_{l}^{m_{j}}(\textbf{\^{k}})]_{\sigma_{1},\sigma_{2}},
\end{eqnarray}
with the elements of the spin-angle matrices
\begin{eqnarray}
[G_{l}^{m_{j}}(\textbf{\^{k}})]_{\sigma_{1},\sigma_{2}}\equiv
\langle\frac{1}{2}\sigma_{1},\frac{1}{2}\sigma_{2}\mid1\sigma_{1}+\sigma_{2}\rangle
\langle1\sigma_{1}+\sigma_{2},lm_{l}\mid1m_{j}\rangle Y_{l}^{m_{l}}(\textbf{\^{k}}),\nonumber\\
\end{eqnarray}
where $m_{j}$ and $m_{l}$ are the projections of the total angular momentum $j=1$ and the orbit angular momentum $l=0,2$
of the pair, respectively. The $Y_{l}^{m_{l}}(\textbf{\^{k}})$ denotes the spherical harmonic with $\textbf{\^{k}}\equiv\textbf{k}/k$.
The anomalous density matrix follows the same expansion. Moreover the
time-reversal invariance implies that
\begin{eqnarray}
\Delta_{\sigma_{1}, \sigma_{2}}(\textbf{k})
=(-1)^{1+\sigma_{1}+\sigma_{2}}\Delta_{-\sigma_{1},-\sigma_{2}}^{*}(\textbf{k}).
\end{eqnarray}
Namely, the pairing gap matrix $\Delta(\textbf{k})$ in spin space
possesses the property
\begin{eqnarray}
\Delta(\textbf{k})\Delta^{\dag}(\textbf{k})=ID^{2}(\textbf{k}),
\end{eqnarray}
i.e., the gap function has the structure of a ``unitary triplet" state \cite{aap}. $I$ is the identity
matrix and $D(\textbf{k})$ is a scalar quantity in spin space.

Once the the isospin singlet $^{3}SD_{1}$ channel has been selected, the pairing gap is an isoscalar and the isospin indices can be dropped off.
The proton/neutron propagators follow from the solution of the
Gorkov equations, and can be present in the form ($\hbar=1$)
\begin{eqnarray}
\textbf{G}_{\sigma,\sigma^{'}}^{(p/n)}(\textbf{k},\omega_{m})=
-\delta_{\sigma,\sigma^{'}}\frac{i\omega_{m}+\xi_{\textbf{k}}\mp \delta\varepsilon_{\textbf{k}}}
{(i\omega_{m}+E_{\textbf{k}}^{+})(i\omega_{m}-E_{\textbf{k}}^{-})}.
\end{eqnarray}
The neutron-proton anomalous propagator matrix in spin space has
the form
\begin{eqnarray}
\textbf{F}^{\dag}(\textbf{k},\omega_{m})=-\frac{\Delta^{\dag}(\textbf{k})}
{(i\omega_{m}+E_{\textbf{k}}^{+})(i\omega_{m}-E_{\textbf{k}}^{-})},
\end{eqnarray}
where $\omega_{m}$ are the Matsubara frequencies, the uper sign in $\textbf{G}_{\sigma,\sigma^{'}}^{(p/n)}$ corresponds to
protons, and the lower to neutrons. The quasiparticle excitation spectra are determined
by finding the poles of the
propagators in Gorkov equations,
\begin{eqnarray}
E_{\textbf{k}}^{\pm}=\sqrt{\xi_{\textbf{k}}^{2}+\frac{1}{2}Tr(\Delta\Delta^{\dag})\pm\frac{1}{2}\sqrt{[Tr(\Delta\Delta^{\dag})]^{2}-
4\det(\Delta\Delta^{\dag})}}\pm \delta\varepsilon_{\textbf{k}},
\end{eqnarray}
where
\begin{eqnarray*}
\xi_{\textbf{k}}\equiv\frac{1}{2}(\varepsilon_{\textbf{k}}^{p}+\varepsilon_{\textbf{k}}^{n}),
\delta\varepsilon_{\textbf{k}}\equiv\frac{1}{2}(\varepsilon_{\textbf{k}}^{p}-\varepsilon_{\textbf{k}}^{n}),
\end{eqnarray*}
and $\varepsilon_{\textbf{k}}^{(n,p)}$ are the single particle energies of neutrons and protons.
Using the ``unitary" property in Eq. (4), the quasiparticle spectra are simplified to
\begin{eqnarray}
E_{\textbf{k}}^{\pm}=\sqrt{\xi_{\textbf{k}}^{2}+D^{2}(\textbf{k})}\pm \delta\varepsilon_{\textbf{k}},
\end{eqnarray}
which are separated into two branches due to the isospin-asymmetry.

In the present ``unitary triplet" case, the gap equation at finite temperature can be written in the standard form
\begin{eqnarray}
\Delta_{\sigma_{1}, \sigma_{2}}(\textbf{k})=-\sum_{\textbf{k}^{'}}\sum_{\sigma_{1}^{'}, \sigma_{2}^{'}}
<\textbf{k}\sigma_{1},-\textbf{k}\sigma_{2}\mid V\mid\textbf{k}^{'}\sigma_{1}^{'},-\textbf{k}^{'}\sigma_{2}^{'}>\nonumber\\
\times\frac{\Delta_{\sigma_{1}^{'}, \sigma_{2}^{'}}(\textbf{k}^{'})}{2\sqrt{\xi_{\textbf{k}^{'}}^{2}+D^{2}(\textbf{k}^{'})}}
[1-f(E_{\textbf{k}^{'}}^{+})-f(E_{\textbf{k}^{'}}^{-})],
\end{eqnarray}
where $f(E)=[1+\exp(\beta E)]^{-1}$ is the Fermi distribution at finite temperature and $V$ is the interaction in the $^{3}SD_{1}$
channel. $\beta^{-1}=k_{B}T$, where $k_{B}$ is the Boltzmann constant
and $T$ is the temperature. Substituting the expansion Eq.(1) into Eqs.(9) and (4), one gets a set of coupled equations for
the quantities $\Delta_{l}^{m_{j}}(k)$
\begin{eqnarray}
\Delta_{l}^{m_{j}}(k)=\frac{-1}{\pi}\int_{0}^{\infty}dk^{'}k^{'2}\sum_{l^{'}=0,2}i^{l^{'}-l}
V^{l^{'}l}_{\lambda}(k^{'},k)\sum_{l^{''}\mu}\Delta_{l^{''}}^{\mu}(k^{'})\nonumber\\
\times\int d\Omega_{\textbf{k}^{'}}Tr[G_{l^{'}}^{m_{j}*}(\textbf{\^{k}}^{'})G_{l^{''}}^{\mu}(\textbf{\^{k}}^{'})]
\frac{1-f(E_{\textbf{k}^{'}}^{+})-f(E_{\textbf{k}^{'}}^{-})}{\sqrt{\xi_{\textbf{k}^{'}}^{2}+D^{2}(\textbf{k}^{'})}},
\end{eqnarray}
with
\begin{eqnarray}
D^{2}(\textbf{k})=\frac{1}{2}Tr(\Delta\Delta^{\dag})=\sum_{ll^{'}=0,2}\sum_{m_{j}m_{j^{'}}}\Delta_{l}^{m_{j}*}(k)
\Delta_{l^{'}}^{m_{j^{'}}}(k)Tr[G_{l}^{m_{j}\dag}(\textbf{\^{k}})G_{l^{'}}^{m_{j^{'}}}(\textbf{\^{k}})],\nonumber\\
\end{eqnarray}
where
\begin{eqnarray}
V^{l^{'}l}_{\lambda}(k^{'},k)\equiv<k^{'}\mid V^{l^{'}l}_{\lambda}\mid k>=\int_{0}^{\infty}r^{2}dr
j_{l^{'}}(k^{'}r)V^{l^{'}l}_{\lambda}(r)j_{l}(k r),
\end{eqnarray}
is the matrix elements of the NN interaction in different partial wave ($\lambda=T,S,l,l^{'}$) channels.
Here $\lambda$ corresponds to the coupled $^{3}SD_{1}$ channel. Following from Eq.(5), we can get the densities of neutrons and protons
\begin{eqnarray}
\rho^{(p/n)}=\sum_{\textbf{k},\sigma}n_{\sigma}^{(p/n)}(\textbf{k}),
\end{eqnarray}
with the distributions
\begin{eqnarray}
n_{\sigma}^{(p/n)}(\textbf{k})=\{\frac{1}{2}(1+\frac{\xi_{\textbf{k}}}{\sqrt{\xi_{\textbf{k}}^{2}+D^{2}(\textbf{k})}})f(E_{\textbf{k}}^{\pm})\nonumber\\
+\frac{1}{2}(1-\frac{\xi_{\textbf{k}}}{\sqrt{\xi_{\textbf{k}}^{2}+D^{2}(\textbf{k})}})[1-f(E_{\textbf{k}}^{\mp})]
\}.
\end{eqnarray}
Summation over frequencies in Eq.(6) leads to the density matrix of the particles in the condensate,
\begin{eqnarray}
\nu(\textbf{k})=\frac{\Delta(\textbf{k})}{2\sqrt{\xi_{\textbf{k}}^{2}+D^{2}(\textbf{k})}}[1-f(E_{\textbf{k}}^{+})-f(E_{\textbf{k}}^{-})].
\end{eqnarray}
It is essential that the coupled Eqs.(10) and (13) should be solved self-consistently.

The six components $\Delta_{l}^{m_{j}}(k)$ of $\Delta(\textbf{k})$ are strongly coupled due to the angle dependent energy denominator
$\sqrt{\xi_{\textbf{k}}^{2}+D^{2}(\textbf{k})}$ in Eqs.(10) and (13). The equations are thus complicated to be solved accurately, and
approximation has been employed. Before introducing the angle-averaging procedure and ADG, we need to substitute $\Delta_{l}^{m_{j}}(k)$ with real
variables. From Eq.(3) we can find the relation
\begin{eqnarray}
\Delta_{l}^{m_{j}*}(k)=-(-1)^{m_{j}}\Delta_{l}^{-m_{j}}(k).
\end{eqnarray}
Therefore, we have four independent components $\Delta_{0}^{0}(k)$, $\Delta_{0}^{1}(k)$, $\Delta_{2}^{0}(k)$ and $\Delta_{2}^{1}(k)$ for $^{3}SD_{1}$ channel
, and we can describe $\Delta_{l}^{m_{j}}(k)$ as
\begin{eqnarray}
\Delta_{0}^{0}(k)&&=i\delta_{0}(k),\nonumber\\
\Delta_{0}^{1}(k)&&=\delta_{1}(k)+i n_{1}(k),\nonumber\\
\Delta_{2}^{0}(k)&&=i\delta_{2}(k),\nonumber\\
\Delta_{2}^{1}(k)&&=\delta_{3}(k)+i n_{3}(k),
\end{eqnarray}
where the six independent variables $\delta_{0}(k)$, $\delta_{1}(k)$, $n_{1}(k)$, $\delta_{2}(k)$, $\delta_{3}(k)$ and $n_{3}(k)$ are
real quantities. Inserting Eq.(17) into Eq.(11), we get
\begin{eqnarray}
D^{2}(\textbf{k})=&&\frac{1}{32\pi}\Big\{4\delta_{0}^{2}(k)-4\sqrt{2}\delta_{0}(k)\delta_{2}(k)[3\cos^{2}\theta-1]
+2\delta_{2}^{2}(k)[3\cos^{2}\theta-1]\nonumber\\
&&+8[\delta_{1}^{2}(k)+n_{1}^{2}(k)]+8[\delta_{3}^{2}(k)+n_{3}^{2}(k)]+6[\delta_{3}^{2}(k)+n_{3}^{2}(k)]\sin^{2}\theta\nonumber\\
&&+4\sqrt{2}n_{1}(k)n_{3}(k)[3\cos^{2}\theta-1]+4\sqrt{2}\delta_{1}(k)\delta_{3}(k)[3\cos^{2}\theta-1]\nonumber\\
&&+12[2\delta_{0}(k)n_{3}(k)+2\delta_{2}(k)n_{1}(k)-\sqrt{2}\delta_{2}(k)n_{3}(k)]\cos \theta\sin \theta\cos\varphi\nonumber\\
&&+12[2\delta_{1}(k)\delta_{2}(k)+2\delta_{0}(k)\delta_{3}(k)-\sqrt{2}\delta_{2}(k)\delta_{3}(k)]\cos \theta\sin \theta\sin \varphi\nonumber\\
&&+6[n_{3}^{2}(k)-\delta_{3}^{2}(k)+2\sqrt{2}\delta_{1}(k)\delta_{3}(k)-2\sqrt{2}n_{1}(k)n_{3}(k)]\sin^{2}\theta\cos 2\varphi\nonumber\\
&&+12[\delta_{3}(k)n_{3}(k)-\sqrt{2}\delta_{1}(k)n_{3}(k)-\sqrt{2}\delta_{3}(k)n_{1}(k)]\sin^{2}\theta\sin 2\varphi
\Big\}.\nonumber\\
\end{eqnarray}

\subsection{The angle-averaging procedure}
Supposing the angle dependence of the energy denominator
$\sqrt{\xi_{\textbf{k}}^{2}+D^{2}(\textbf{k})}$ can be neglected, the gap equations are simplified by substituting $D^{2}(\textbf{k})$ with its
angular average value,
\begin{eqnarray}
D^{2}(\textbf{k})\rightarrow &&d^{2}(k)=\frac{1}{4\pi}\int d\Omega_{\textbf{k}}D^{2}(\textbf{k})\nonumber\\
&&=\frac{1}{8\pi}\Big[2\delta_{1}^{2}(k)+
\delta_{0}^{2}(k)+2n_{1}^{2}(k)+2\delta_{3}^{2}(k)+
\delta_{2}^{2}(k)+2n_{3}^{2}(k)
\Big].
\end{eqnarray}
Thereby, the energy denominator and the quasiparticle spectra are isotropic. Noting the properties of $G_{l}^{m_{j}}(\textbf{\^{k}})$
\begin{eqnarray}
\int d\Omega_{\textbf{k}}Tr[G_{l}^{m_{j}*}(\textbf{\^{k}})G_{l^{'}}^{m_{j^{'}}*}(\textbf{\^{k}})]=\delta_{ll^{'}}\delta_{m_{j}m_{j^{'}}},
\end{eqnarray}
the different $m_{j}$ components $\Delta_{l}^{m_{j}}(k)$ with the same $l$ become uncoupled and all equal to each other. It follows that
\begin{eqnarray}
\delta_{1}(k)=n_{1}(k)=\sqrt{\frac{1}{2}}\delta_{0}(k), \delta_{3}(k)=n_{3}(k)=\sqrt{\frac{1}{2}}\delta_{2}(k),
\end{eqnarray}
and
\begin{eqnarray}
d^{2}(k)=\frac{3}{8\pi}[\delta_{0}^{2}(k)+\delta_{2}^{2}(k)].
\end{eqnarray}
Taking the normalization
\begin{eqnarray}
\Delta_{0}(k)=\sqrt{\frac{3}{8\pi}}\delta_{0}(k),\Delta_{2}(k)=-\sqrt{\frac{3}{8\pi}}\delta_{2}(k),
\end{eqnarray}
the set of equations in Eq.(10) reduces to two coupled equations for the $^{3}S_{1}$ and $^{3}D_{1}$ gap components
$\Delta_{0}(k)$ and $\Delta_{2}(k)$, respectively. They read
\begin{eqnarray}
\left(
\begin{array}{l}
\Delta_{0} \\
\Delta_{2}
\end{array}
\right)(k)=\frac{-1}{\pi}\int dk^{'}k^{'2}\left(
\begin{array}{ll}
V^{00} & V^{02}\\
V^{20} & V^{22}
\end{array}
\right)(k,k^{'})\frac{1-f(E_{k^{'}}^{+})-f(E_{k^{'}}^{-})}{\sqrt{\xi_{\textbf{k}^{'}}^{2}+D^{2}(k^{'})}}\left(
\begin{array}{l}
\Delta_{0} \\
\Delta_{2}
\end{array}\right)(k^{'})\nonumber\\,
\end{eqnarray}
where $V^{00}$, $V^{02}$, $V^{20}$, $V^{22}$ are given in Eq.(12) with $l,l^{'}=0,2$ and
\begin{eqnarray}
E^{\pm}_k=\sqrt{\xi_{\textbf{k}}^{2}+D^{2}(k)}\pm\delta\varepsilon_{\textbf{k}}, &D^{2}(k)\equiv d^{2}(k)=\Delta_{0}^{2}(k)+\Delta_{2}^{2}(k).
\end{eqnarray}
Eqs.(13), (24) and (25) compose the angle-averaged gap equations
and should be solved simultaneously for isospin-asymmetric
nuclear matter. The quasiparticle spectra here are isotropic and
the gapless excitation exists at large asymmetry
($|\delta\varepsilon_{\textbf{k}_{F}}|\geq D(k_{F})$) near zero
temperature.

\subsection{The angle dependent gap}
As pointed out in the Sec.I, the angle dependence of quasiparticle
spectra due to $D^{2}(\textbf{k})$ may increase the phase-space
overlap of neutron and proton near their average Fermi surface. We
consider an axi-symmetric $D^{2}(\textbf{k})$ solution which
corresponds to an axi-symmetric deformation of the neutron and
proton Fermi spheres. From the expression in Eq.(18), the
axi-symmetric solutions are restricted by
\begin{eqnarray}
2\delta_{0}(k)n_{3}(k)+2\delta_{2}(k)n_{1}(k)-\sqrt{2}\delta_{2}(k)n_{3}(k)=0,\nonumber\\
2\delta_{1}(k)\delta_{2}(k)+2\delta_{0}(k)\delta_{3}(k)-\sqrt{2}\delta_{2}(k)\delta_{3}(k)=0,\nonumber\\
n_{3}^{2}(k)-\delta_{3}^{2}(k)+2\sqrt{2}\delta_{1}(k)\delta_{3}(k)-2\sqrt{2}n_{1}(k)n_{3}(k)=0,\nonumber\\
\delta_{3}(k)n_{3}(k)-\sqrt{2}\delta_{1}(k)n_{3}(k)-\sqrt{2}\delta_{3}(k)n_{1}(k)=0.
\end{eqnarray}
There exists only one nontrivial solution
\begin{eqnarray}
\delta_{1}(k)=n_{1}(k)=\delta_{3}(k)=n_{3}(k)=0,
\end{eqnarray}
which corresponds to the $m_{j}=0$ gap components of $\Delta_{l}^{m_{j}}(k)$. In this case
\begin{eqnarray}
D^{2}(\textbf{k})\rightarrow D^{2}(k,\theta)=&&\frac{1}{8\pi}\Big[\delta_{0}^{2}(k)
-\sqrt{2}\delta_{0}(k)\delta_{2}(k)(3\cos^{2}\theta-1)\nonumber\\
&&+\delta_{2}^{2}(k)\frac{3\cos^{2}\theta+1}{2}
\Big].
\end{eqnarray}
Using the normalization
\begin{eqnarray}
\Delta_{0}(k)=\sqrt{\frac{1}{8\pi}}\delta_{0}(k),\Delta_{2}(k)=-\sqrt{\frac{1}{8\pi}}\delta_{2}(k),
\end{eqnarray}
one gets the angle dependent gap equations
\begin{eqnarray}
\left(
\begin{array}{l}
\Delta_{0} \\
\Delta_{2}
\end{array}
\right)(k)&&=\frac{-1}{\pi}\int dk^{'}k^{'2}\left(
\begin{array}{ll}
V^{00} & V^{02}\\
V^{20} & V^{22}
\end{array}
\right)(k,k^{'})\nonumber\\
&&\times\int d\Omega_{\textbf{k}^{'}}\frac{1-f(E_{k^{'}}^{+})-f(E_{k^{'}}^{-})}{\sqrt{\xi_{\textbf{k}^{'}}^{2}+D^{2}(k^{'},\theta)}}
\left(\begin{array}{ll}
\emph{f}(\theta) & \emph{g}(\theta)\\
\emph{g}(\theta) & \emph{h}(\theta)
\end{array}
\right)
\left(
\begin{array}{l}
\Delta_{0} \\
\Delta_{2}
\end{array}\right)(k^{'}),
\end{eqnarray}
with the following axi-symmetric quantities,
\begin{eqnarray}
D^{2}(k,\theta)&&=\Delta_{0}^{2}(k)+\sqrt{2}\Delta_{0}(k)\Delta_{2}(k)[3\cos^{2}\theta-1]+\Delta_{2}^{2}(k)
[\frac{3\cos^{2}\theta+1}{2}], \nonumber\\ E^{\pm}_k &&=\sqrt{\xi_{\textbf{k}}^{2}+D^{2}(k,\theta)}\pm\delta\varepsilon_{\textbf{k}}.
\end{eqnarray}
The angle matrix $\left(\begin{array}{ll}
\emph{f}(\theta) & \emph{g}(\theta)\\
\emph{g}(\theta) & \emph{h}(\theta)
\end{array}
\right)$ comes from the coupling among the different $m_{j}$ components of $\Delta(\textbf{k})$. The matrix elements are
\begin{eqnarray}
\emph{f}(\theta)&&=Tr[G_{0}^{0*}(\textbf{\^{k}}^{'})G_{0}^{0}(\textbf{\^{k}}^{'})]=\frac{1}{4\pi},\nonumber\\
\emph{g}(\theta)&&=-Tr[G_{0}^{0*}(\textbf{\^{k}}^{'})G_{2}^{0}(\textbf{\^{k}}^{'})]=\frac{\sqrt{2}}{8\pi}(3\cos^{2}\theta-1),\nonumber\\
\emph{h}(\theta)&&=Tr[G_{2}^{0*}(\textbf{\^{k}}^{'})G_{2}^{0}(\textbf{\^{k}}^{'})]=\frac{1}{8\pi}(3\cos^{2}\theta+1).
\end{eqnarray}

As a first inspection, when applying following the substitution [both in the
gap equations (30) and the expression of $E_{k}^{\pm}$ in Eq.(31)]
\begin{eqnarray}
\frac{3\cos^{2}\theta}{8\pi}\rightarrow\frac{1}{8\pi},
\end{eqnarray}
which has been used as the angle-averaging procedure for
$^{3}PF_{2}$ superfluidity in Ref.\cite{3pf2}, Eq.(30) reduces to
the form of angle-averaged gap Eq.(24). At zero temperature, the
pairing is suppressed by the gapless excitation near the average Fermi
surface in the AAG state. However, pairing can exist
in the interval $(0,\theta_{1})\bigcup(\pi,\pi-\theta_{1})$ of
$\theta$ near the average Fermi surface in the ADG state, where
\begin{eqnarray*}
\cos^{2}\theta_{1}=\frac{\delta\mu^{2}-\Delta_{0}^{2}(k_{F})+\sqrt{2}\Delta_{0}(k_{F})\Delta_{2}(k_{F})-\Delta_{2}^{2}(k_{F})/2}
{3\sqrt{2}\Delta_{0}(k_{F})\Delta_{2}(k_{F})+3\Delta_{2}^{2}(k_{F})/2}
\end{eqnarray*}
and $\delta\mu$ is the difference between the neutron and proton chemical potentials. This mechanism is consistent with that of the FFLO state.
Furthermore, the influences from the coupling of different $m_{j}$ components are partially taken into account via the angle matrix $\left(\begin{array}{ll}
\emph{f}(\theta) & \emph{g}(\theta)\\
\emph{g}(\theta) & \emph{h}(\theta)
\end{array}
\right)$ in the ADG state.

\subsection{Thermodynamics}
For isospin-asymmetric nuclear matter at a fixed temperature and given neutron and proton densities,
the essential quantity to describe the thermodynamics of the system is the free energy defined as
\begin{eqnarray}
\emph{F}|_{\rho,\beta}=\emph{U}-\beta^{-1}\emph{S},
\end{eqnarray}
where $\emph{U}$ is the internal energy and $\emph{S}$ is the entropy. In the mean-field approximation,
the entropy of the superfluid state is
\begin{eqnarray}
\emph{S}=-2k_{B}\sum_{\textbf{k}}&&\{f(E_{\textbf{k}}^{+})\ln f(E_{\textbf{k}}^{+})+\bar{f}(E_{\textbf{k}}^{+})\ln \bar{f}(E_{\textbf{k}}^{+})\nonumber\\
&&+f(E_{\textbf{k}}^{-})\ln f(E_{\textbf{k}}^{-})+\bar{f}(E_{\textbf{k}}^{-})\ln \bar{f}(E_{\textbf{k}}^{-})
\},
\end{eqnarray}
where $\bar{f}(E_{\textbf{k}}^{\pm})=1-f(E_{\textbf{k}}^{\pm})$. The internal energy of the superfluid state reads
\begin{eqnarray}
\emph{U}&&=\sum_{\sigma\textbf{k}}[\varepsilon_{\textbf{k}}^{(n)}n_{\sigma}^{(n)}(\textbf{k})+\varepsilon_{\textbf{k}}^{(p)}n_{\sigma}^{(p)}(\textbf{k})]\nonumber\\
&&+\sum_{\textbf{k},\textbf{k}^{'}}\sum_{\sigma_{1},\sigma_{2},\sigma_{1}^{'},\sigma_{2}^{'}}
<\textbf{k}\sigma_{1},-\textbf{k}\sigma_{2}\mid V\mid\textbf{k}^{'}\sigma_{1}^{'},-\textbf{k}^{'}\sigma_{2}^{'}>
\nu^{\dag}_{\sigma_{2},\sigma_{1}}(\textbf{k})
\nu_{\sigma_{1}^{'},\sigma_{2}^{'}}(\textbf{k}^{'}).
\end{eqnarray}
The first term in Eq.(36) includes the kinetic energy of the quasiparticles which is a functional of the pairing gap. In the
normal state it reduces to the kinetic energy of the neutrons and protons. The second term includes the BCS mean-field interaction among
the particles in the condensate and can be eliminated in terms of the gap equation (9) (shown in Appendix). Finally, the internal energy is written as
\begin{eqnarray}
\emph{U}&&=\sum_{\sigma\textbf{k}}[\varepsilon_{\textbf{k}}^{(n)}n_{\sigma}^{(n)}(\textbf{k})+\varepsilon_{\textbf{k}}^{(p)}n_{\sigma}^{(p)}(\textbf{k})]\nonumber\\
&&-\sum_{\textbf{k}}\frac{D^{2}(\textbf{k})}{\sqrt{\xi_{\textbf{k}}^{2}+D^{2}(\textbf{k})}}[1-f(E_{\textbf{k}}^{+})-f(E_{\textbf{k}}^{-})].
\end{eqnarray}
A thermodynamically stable state minimizes the difference of the free energies
between the superconducting and normal states, $\delta\emph{F}=\emph{F}_{S}-\emph{F}_{N}$ [the
free energy in the normal state follows from Eqs.(35) and (37) when $\Delta\rightarrow0$].


\section{Results}
The numerical calculations here focus on the effects of the
angle dependence of the quasiparticle spectra and the emergence of
the ADG phase in isospin-asymmetric nuclear matter. To simplify
the calculations, several assumptions have been adopted. Firstly,
the pairing interaction is approximated by the bare interaction;
i.e., the effects of the screening of the pairing interaction are
ignored. 
Secondly, we adopt the free single particle (s.p.) spectrum, which may
affect the density of the states at the Fermi surface.
Previous calculations \cite{bhf1,bhf2} show that using a more
realistic s.p. spectrum obtained from the BHF approach (the BHF
spectrum) may reduce the $^{3}SD_{1}$ channel pairing gap as
compared with the free spectrum. As for the pairing interaction, the
screening potential (i.e., the higher-order contribution in the
pairing interaction) for the $^{3}S_{1}$ pairing channel in nuclear
matter under different approximations has been discussed in
Refs.\cite{scr1}. It has been shown the screening potential is
repulsive at low densities in the one-bubble approximation,
whereas it is slightly attractive in the full RPA (suitably
renormalized to cure the low density mechanical instability of
nuclear matter \cite{scr1,scr2}). Up to now, the screening effect
on the pairing gap remains an open problem. Finally, we ignore the
isospin triplet states, which is valid when the pairing in the
isospin singlet channel is much larger than that in the isospin
triplet channel. However, the argument could be questionable when the
first two approximations are abandoned. In the present
calculations, the net density is fixed at the
empirical saturation density of nuclear matter
$\rho=\rho_{0}=0.17fm^{-3}$ except for Fig.7, and the Argonne $V_{18}$ potential is adopted as the pairing interaction.



\begin{figure}
\includegraphics[scale=0.7]{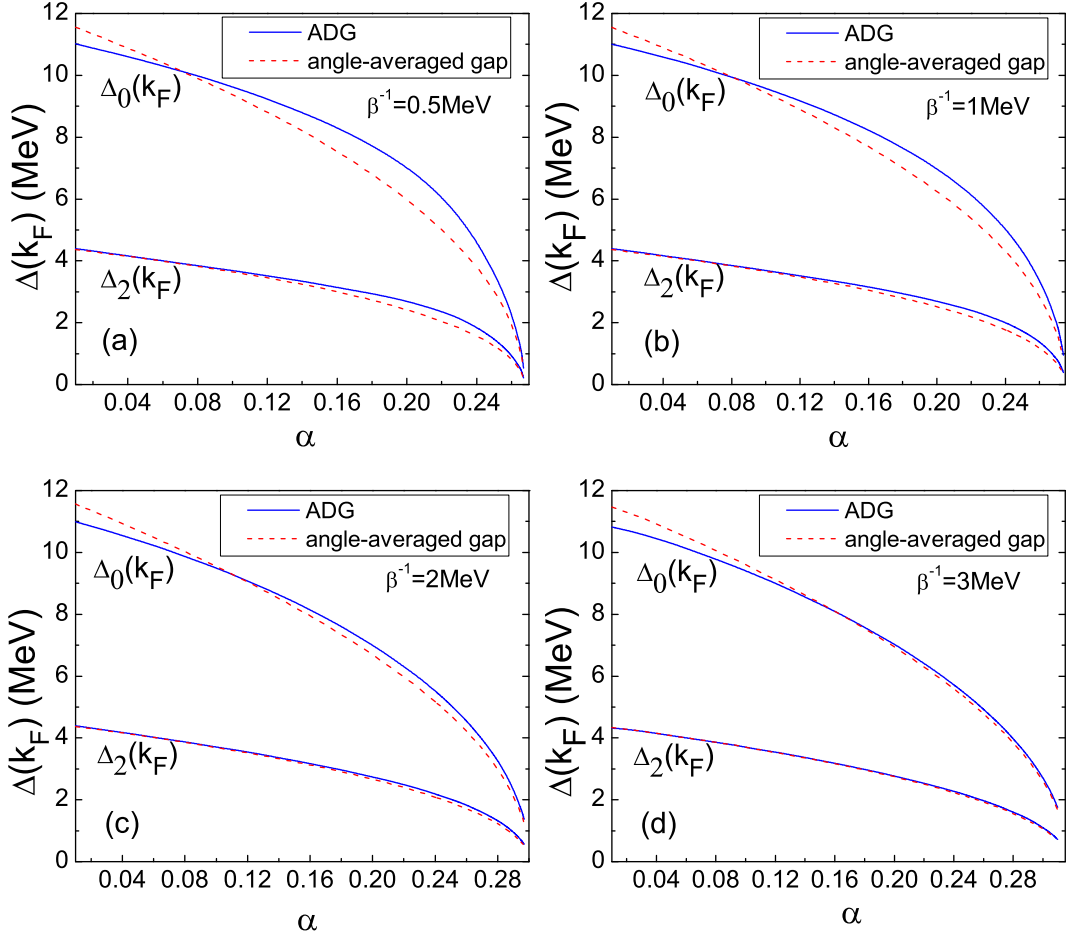} \caption{(Color online). The upper and lower curves in the figures are related to the
values of $\Delta_{0}(k_{F})$ and $\Delta_{2}(k_{F})$ vs
isospin-asymmetry $\alpha$. The blue solid and red dashed lines
correspond to the ADG and angle-averaged gap, respectively. }
\label{gap}
\end{figure}


Fig.1 shows the angle-averaged and angle dependent gaps
$\Delta_{0}(k_{F})$ and $\Delta_{2}(k_{F})$ in the $^{3}SD_{1}$
partial-wave channel as a function of isospin-asymmetry $\alpha$, defined
as $\alpha=(\rho_{n}-\rho_{p})/\rho$. The temperatures are
set at low-temperature regime $\beta^{-1}=$ $0.5$ MeV, $1.0$ MeV,
$2.0$ MeV, $3.0$ MeV (the critical temperature $\beta^{-1}_{c}$
where the superfluid vanishes is about $7.5$ MeV for isospin-symmetric
case).
 At temperature $\beta^{-1}= 0.5$ MeV, the value of $\Delta_{0}(k_{F})$ in the ADG state becomes larger than
that of the angle-averaged gap state for $\alpha\geq0.07$, and the difference of $\Delta_{0}(k_{F})$ between the ADG and angle-averaged gap states reaches 22 percent at $\alpha=0.23$. With increasing temperature, the difference of $\Delta_{0}(k_{F})$ between the two kinds of states decreases rapidly.
 The critical
isospin-asymmetries $\alpha_{c}$ at which the gaps vanish are the
same in the two states, and their values are $0.267$, $0.275$, $0.30$ and
$0.315$ for the temperatures $0.5$ MeV, $1.0$ MeV, $2.0$ MeV and
$3.0$ MeV, respectively. It implies that the thermal excitation can promote pairing in large isospin-asymmetry nuclear matter at low temperature regime.

In order to have an entire inspection of the difference between the pairing gaps of the ADG state and the angle-averaged gap state,
we exhibit the gap functions in Fig.2.
\begin{figure}
\includegraphics[scale=0.7]{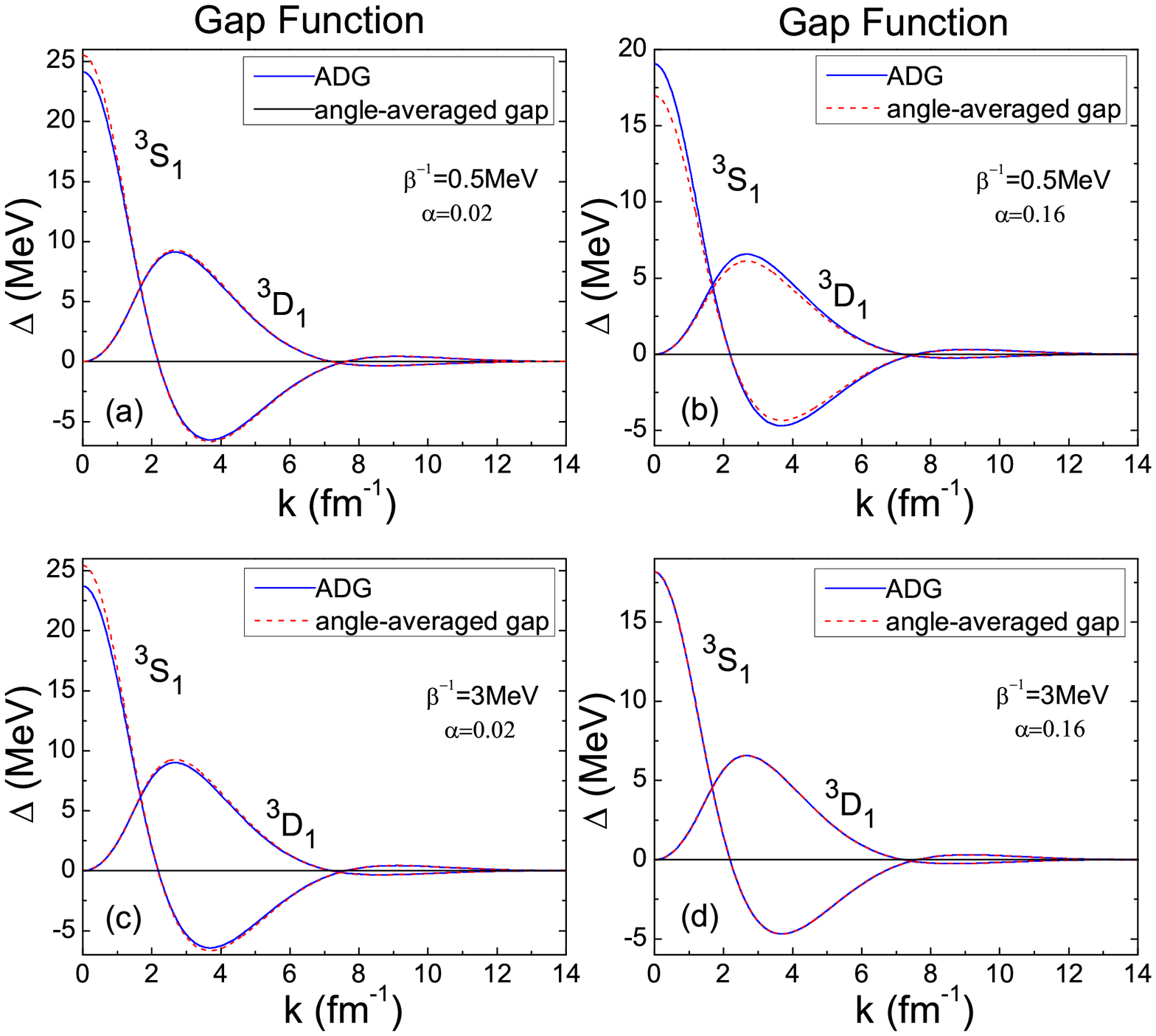} \caption{(Color online). The curves marked with symbols $^{3}S_{1}$ and $^{3}D_{1}$ are related to the gap functions $\Delta_{0}(k)$ and $\Delta_{2}(k)$ in Eqs.(24) and (30). The blue solid and red dashed lines correspond to the ADG and angle-averaged gap, respectively.} \label{wave}
\end{figure}
At temperature $\beta^{-1}=0.5$ MeV, the gap functions of the two different kinds of states are almost the same except a little
difference of $\Delta_{0}(k)$ near the zero momentum
for the asymmetry $\alpha=0.02$ [in Fig.2.(a)]. When the system
becomes more asymmetric, the difference gets larger [in
Fig.2.(b)]. However, the curves of the ADG coincide with these of the
angle-averaged gap for $\beta^{-1}=3.0$ MeV with $\alpha=0.16$ [in Fig.2.(d)].
That implies the angle-averaging procedure is a satisfactory
approximation for asymmetric nuclear matter at high temperatures.

\begin{figure}
\includegraphics[scale=0.7]{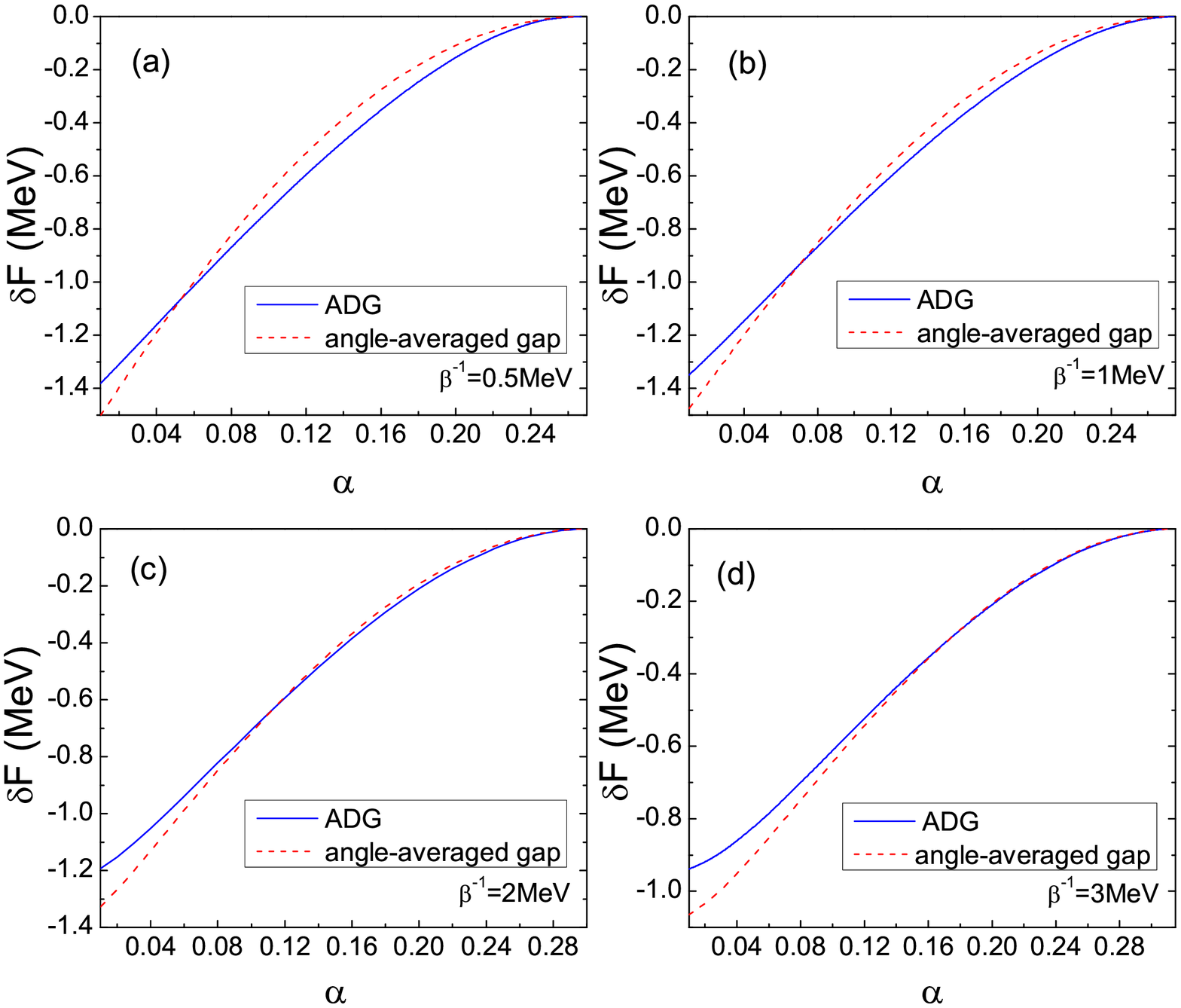} \caption{(Color online). The difference of the free energy between the superconducting and normal states as a function of the isospin-symmetry $\alpha$ for diffderent temperatures.
The blue solid and red dashed lines correspond to the ADG and
angle-averaged gap, respectively.} \label{dife}
\end{figure}
A larger gap value in ADG state may result in a larger pairing energy in the condensate [second term in Eqs.(36) and
(37)], which has important influence on the free energy of the
superconducting state. Thus we calculate the free-energy difference $\delta\emph{F}$ between the normal and superconducting states.
 The results are shown in Fig.3, where the
parameters are set as the same as those in Fig.1. At
temperature $\beta^{-1}=0.5$ MeV, $\delta F$ in the ADG state gets
smaller than that of the angle-averaged gap state when $\alpha\geq0.06$,
especially, the former is about 35 percent lower than the latter
in the regime $\alpha>0.17$. We can conclude that the ADG state is more
favored than the angle-averaged gap state for large asymmetry at
low temperature, since the angle dependence of the pairing gap enhances the pairing
energy and has little effect on the kinetic energy. However, the
thermal excitation can reduce the effects of angle dependence of the pairing gap [comparing the Fig.3.(a) with Fig.3.(d)].
It is also shown in Fig.3 that the values of $\delta F$ tend to zero gently
when $\alpha\rightarrow\alpha_{c}$ at
different temperatures.

\begin{figure}
\includegraphics[scale=0.7]{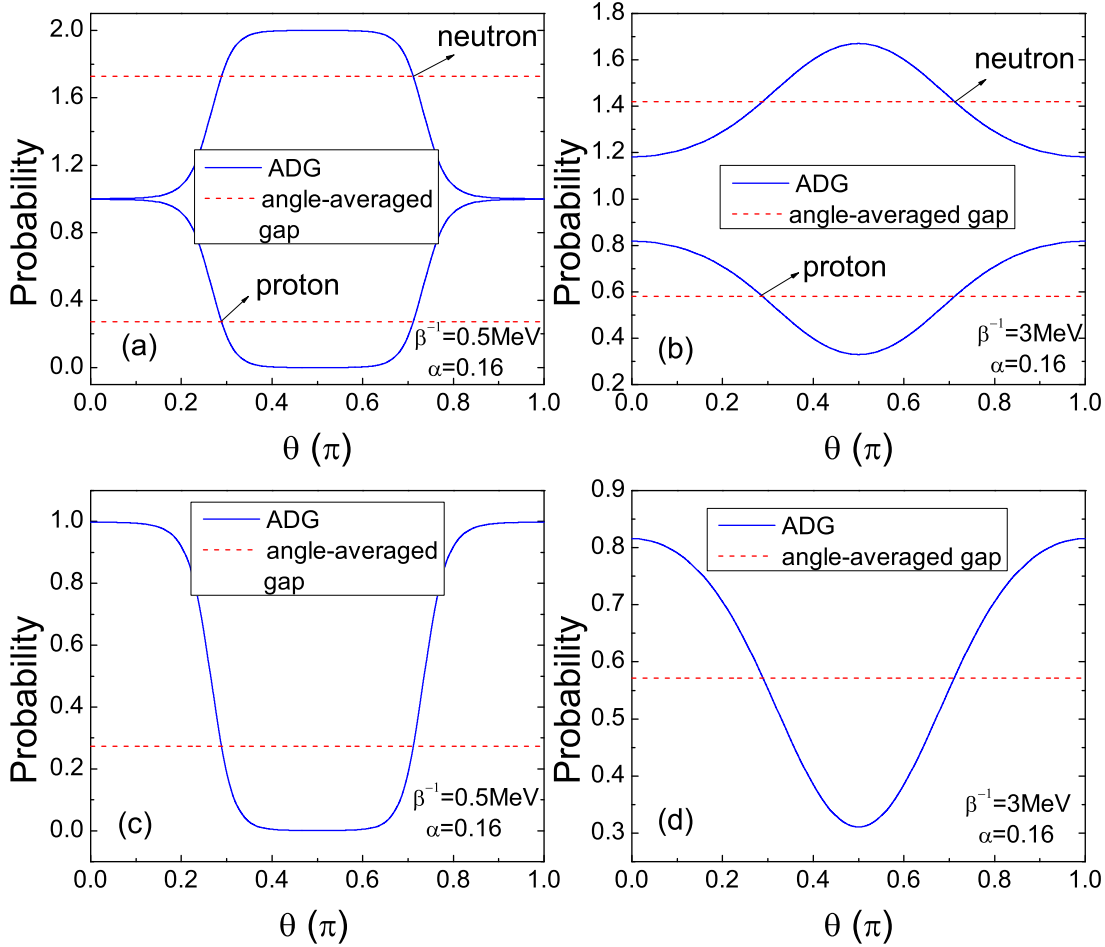} \caption{(Color online). The higher and lower curves in the upper two figures are related to the neutron and proton occupation probabilities, respectively. The curves in the lower two figures are the pairing probabilities.
The blue solid and red dashed lines correspond to the ADG and
angle-averaged gap, respectively.} \label{pro}
\end{figure}
One straightforward way to understand the effects of angle
dependence of the pairing gap is to investigate the normal and superconducting
occupation probabilities [obtained from Eqs.(14) and (15)]
near the average Fermi surface (related to the average chemical
potential of neutron and proton). The results are depicted in
Fig.4, where the spin summation has been carried out. In this figure, the neutron/proton and pairing particle occupation probabilities
at the average Fermi surface for a fixed asymmetry $\alpha=0.16$ at temperature $\beta^{-1}=$ $0.5$ MeV has been compared with those at $3.0$ MeV.
In
isospin-asymmetric nuclear matter, the large splitting between the neutron and proton
occupation probabilities prevents the pairing around the average
Fermi surface in the angle-averaging procedure. However, in the
ADG state, the splitting is reduced by the angle dependence of the pairing gap in partial area around
the average Fermi surface, i.e., in the regime
$\theta\subset(0,\frac{\pi}{5})\cup(\frac{4\pi}{5},\pi)$ as shown
in Fig.4.(a). In Fig.4.(c), as compared with the angle-averaged
gap, although the pairing in the ADG state is almost fully
suppressed in the regime
$\theta\subset(\frac{\pi}{5},\frac{4\pi}{5})$ in ADG state, it is
obviously enhanced at $\theta$ smaller than $\frac{\pi}{5}$ and
greater than $\frac{4\pi}{5}$.

Substituting the expression of $D^{2}(\textbf{k})$ in Eq.(31) into Eq.(14), we can find that the Fermi spheres of neutron and proton are no longer isotropic in the ADG state. Since we assume an axi-symmetric quasiparticle spectrum in the ADG state, the rotational symmetry is spontaneously broken [in terms of group theory the $O(3)$ symmetry breaks down to $O(2)$] and there exists one favored direction.
The neutron Fermi sphere possesses an oblate deformation perpendicular to the favored direction, whereas the proton Fermi sphere has a prolate
deformation along the favored direction. The two different deformations enhance the correlation between neutrons and protons near their average Fermi surface. 
However, at high temperature the neutron/proton occupation
probability in the ADG becomes almost isotropic as shown in Fig.4.(b),
namely, the thermal excitation reduces the angle dependence of
quasiparticle spectra. In this case, the deformation of the
neutron/proton Fermi sphere fails to increase the phase-space
overlap of neutron and proton near their average Fermi surface
effectively. Thus the results of the ADG state are nearly the same as
that of the angle-averaged gap state, i.e., the angle-averaging
procedure becomes an adequate approximation at high temperatures
$\beta^{-1}\geq3 $ MeV.

\begin{figure}
\includegraphics[scale=0.7]{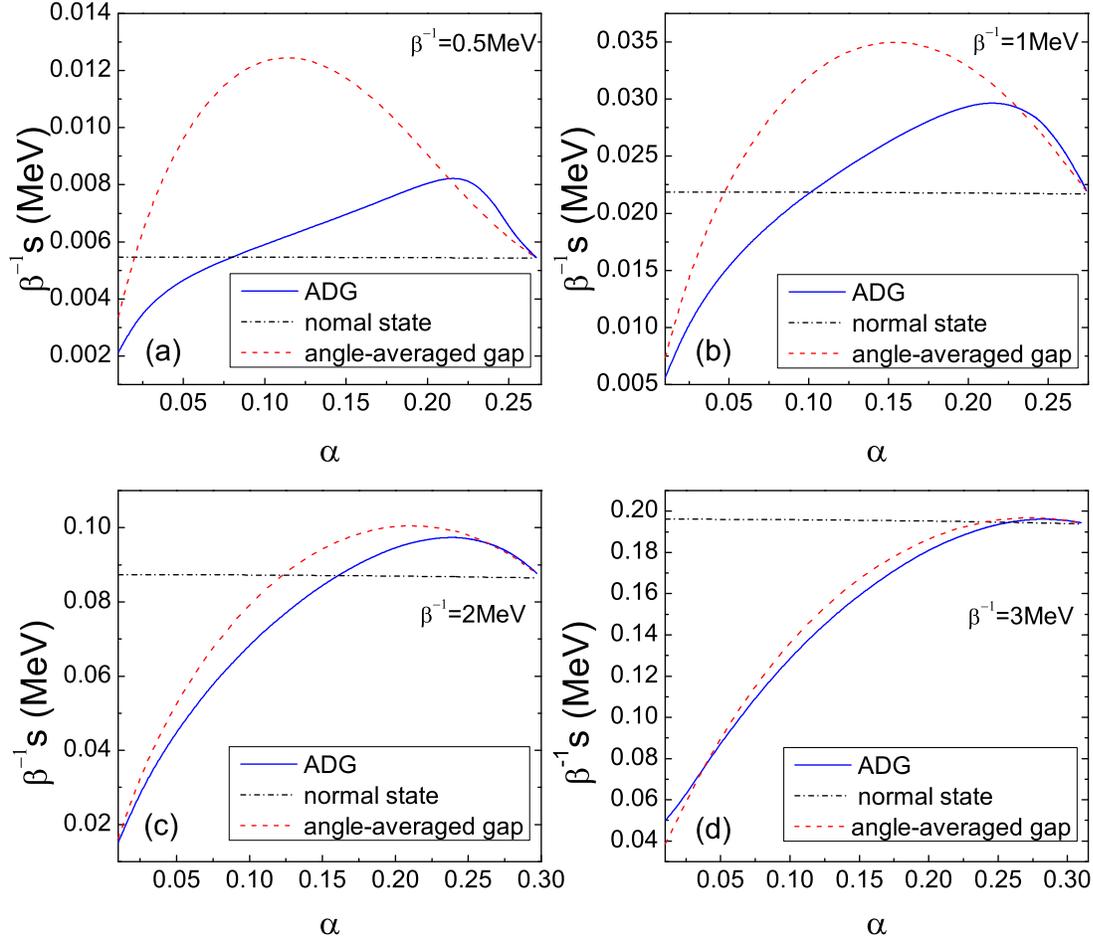} \caption{(Color online). The entropy (scaled by $\beta^{-1}$) as a function of the isospin-symmetry $\alpha$ for different temperature. The blue solid, red dashed and black dash-dotted lines correspond to the ADG, angle-averaged gap and normal state, respectively.} \label{sen}
\end{figure}
Fig.5 displays the entropy ($\beta^{-1}S$) as a function of
isospin-asymmetry $\alpha$ for different temperatures
$\beta^{-1}=$ $0.5$ MeV, $1.0$ MeV, $2.0$ MeV and $3.0$ MeV.  The
entropy in the superconducting state is smaller than that in
the normal state near $\alpha=0$, and gets larger than that in the
normal state at sufficiently large asymmetry. However, around the
transition point $\alpha_{c}$ from the superconducting state to the normal state, the
entropies of the superconducting states (both of the ADG state and the angle-averaged gap state) approach to the value of the
normal state, i.e., the latent heats
$Q=\beta^{-1}(S_{s}-S_{n})\rightarrow0$ when $\alpha\rightarrow\alpha_{c}$. Hence
the transitions are of second order. At temperature
$\beta^{-1}=$ $0.5$ MeV, the entropy in the ADG state is nearly a
linear function of the isospin-asymmetry when $0.02<\alpha<0.22$. With increasing temperature, the linear property of the entropy curve
disappears and the difference between the ADG state and the angle-averaged
gap state gets smaller.

\begin{figure}
\includegraphics[scale=0.7]{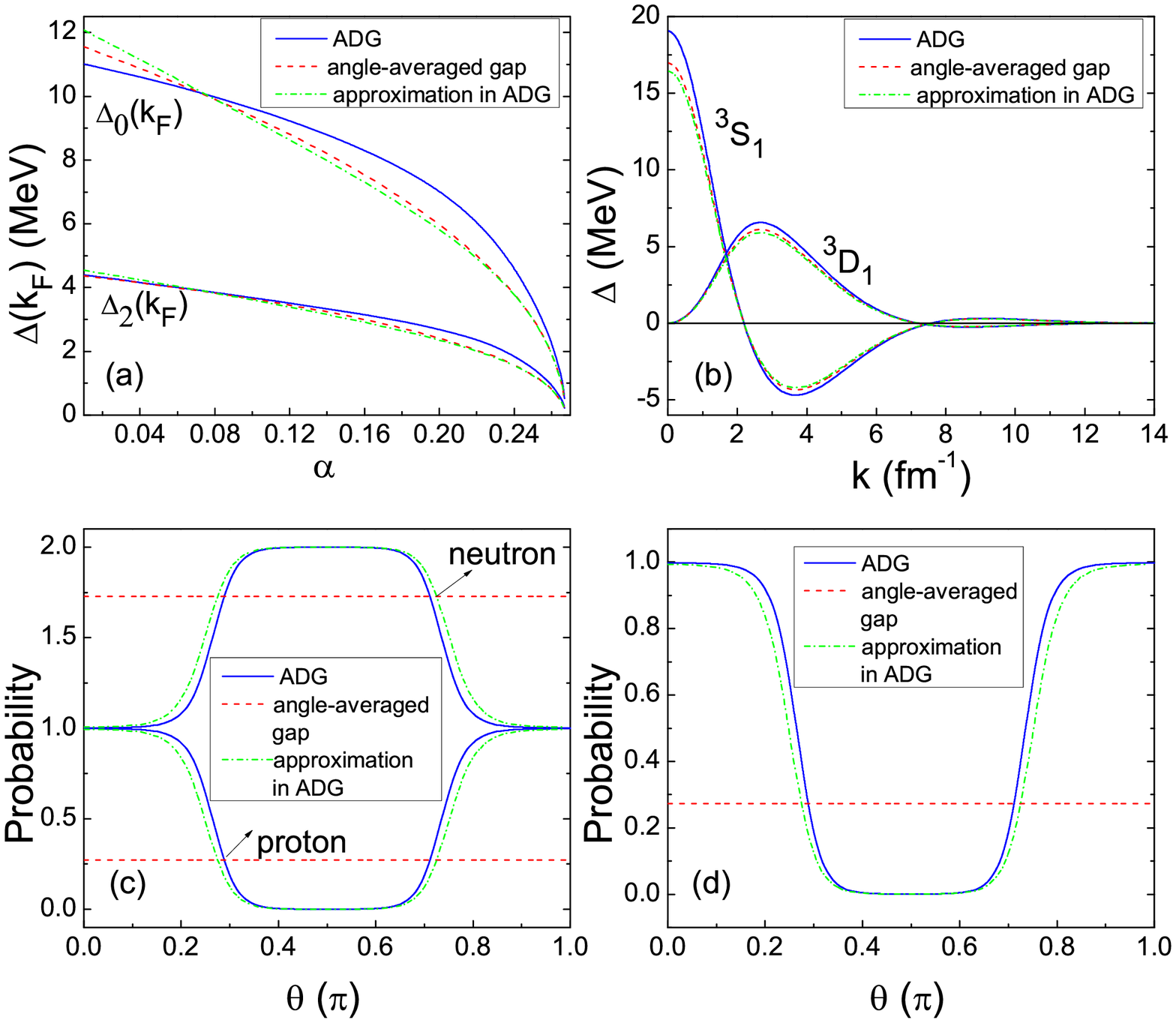} \caption{(Color online). $\Delta_{0}(k_{F})$ and $\Delta_{2}(k_{F})$ as a function of isospin-asymmetry $\alpha$ for the ADG, angle-averaged gap and `approximation in ADG' are shown in Fig.(a). Fig.(b) exhibits the gap functions for the three case. The normal and superconducting occupation probabilities at the average Fermi surface for the three case are shown in Figs.(c) and (d), respectively. The blue solid, red dashed and green dash-dotted lines correspond to the ADG, angle-averaged gap and the `approximation in ADG', respectively. The temperature is set to be $\beta^{-1}=0.5$ MeV, and the isospin-asymmetry $\alpha=0.16$ in (b), (c), (d).} \label{2ap}
\end{figure}
Comparing the gap equations (30) for the ADG state with Eq.(24) for the angle-averaged gap state, two differences appear in the ADG state, i.e., the angle dependent
quasiparticle spectrum and the angle matrix $\left(\begin{array}{ll}
\emph{f}(\theta) & \emph{g}(\theta)\\
\emph{g}(\theta) & \emph{h}(\theta)
\end{array}
\right)$. The first leads to the deformation of neutron/proton
Fermi sphere, and the second corresponds to the coupling among
different $m_{j}$ gap components. Actually, the angle matrix
modifies the strength of $V^{l^{'}l}_{\lambda}(k^{'},k)$ in
different directions in momentum space. We replace the angle
matrix by $\frac{1}{4\pi}\left(\begin{array}{ll}
1 & 0\\
0 & 1
\end{array}
\right)$ to inspect the influence of the angle matrix. The results
are shown in Fig.6 for asymmetry $\alpha=0.16$ in (b), (c), (d)
and the temperature is set to be $\beta^{-1}=0.5$ MeV. The dash-doted lines denoted by `approximation in ADG' are obtained by
replacing the angle matrix with
$\frac{1}{4\pi}\left(\begin{array}{ll}
1 & 0\\
0 & 1
\end{array}
\right)$. Figs.6.(c) and (d) exhibit the neutron/proton and
pairing particle occupation probabilities at the average Fermi
surface, respectively. The curves of ADG and `approximation in ADG' are nearly
the same in Figs.6.(c) and (d). Whereas the gap functions in
Fig.6.(b) show that the curves of `approximation in ADG' behave
closer to those of the angle-averaged gap state than those of the ADG state. Fig.6.(a) displays
the $\Delta_{0}(k_{F})$ and $\Delta_{2}(k_{F})$ vs isospin-asymmetry
$\alpha$. The gaps of `approximation in ADG' turn out to be
smaller than both the gaps in the ADG state and angle-averaged gap state
when $\alpha>0.07$.
 Moreover, the curves of `approximation in ADG' are much closer to
that of the angle-averaged gap state. All these results indicate that
the influence of the angle matrix is much more important than that
of the angle dependence of quasiparticle spectrum. Furthermore,
the coupling from
different $m_{j}$ gap components may strengthen the pairing interaction for large isospin-asymmetry
at low temperatures.

\begin{figure}
\includegraphics[scale=0.7]{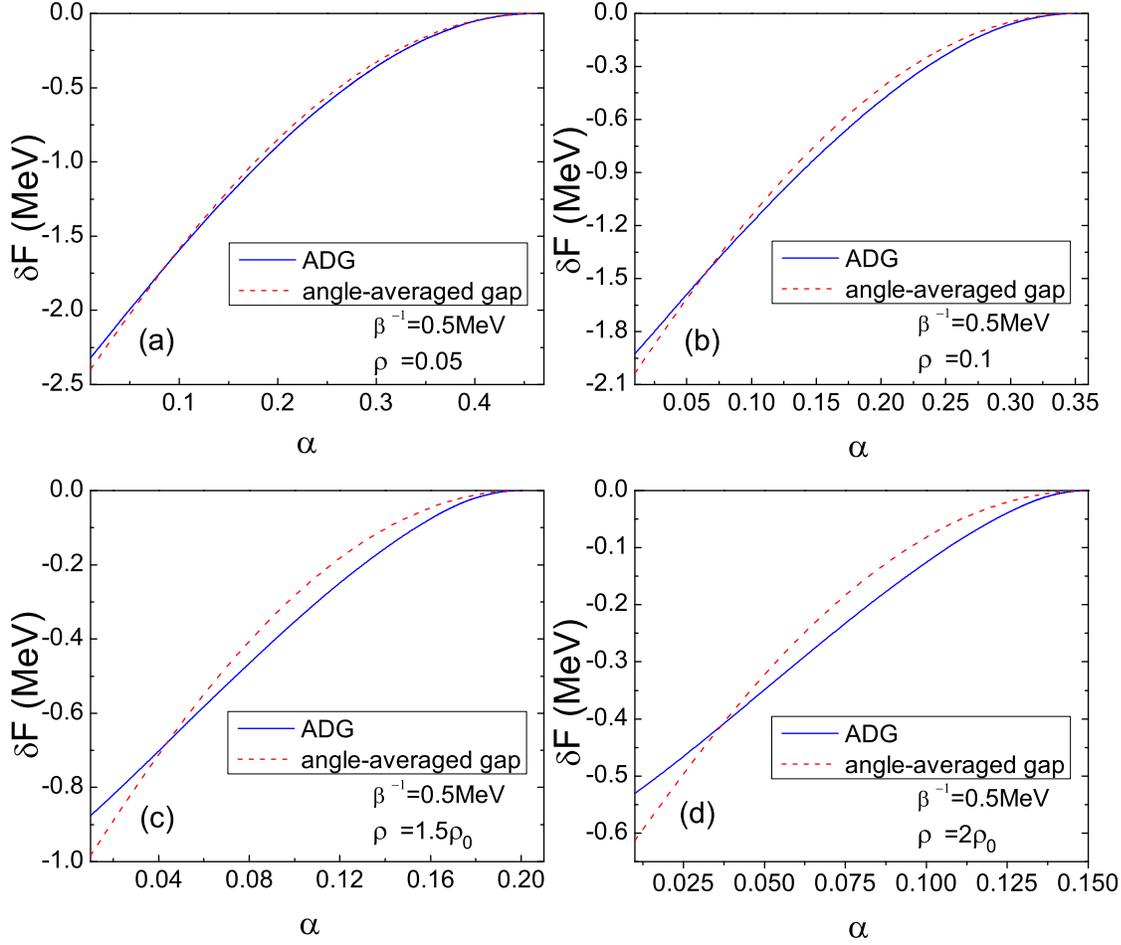} \caption{(Color online). The difference of the free energy between the superconducting and normal states as a function of the isospin-symmetry $\alpha$ for different densities at a fixed temperature $\beta^{-1}=0.5$ MeV. The blue solid and red dashed lines correspond to the ADG and angle-averaged gap, respectively.} \label{difr}
\end{figure}
In order to discuss the effect of angle dependence of the
pairing gap for different densities, we show the free energy difference $\delta\emph{F}$
between the superconducting and normal states at temperature $\beta^{-1}=0.5$ MeV vs isospin-asymmetry $\alpha$ in Fig.7.
 The densities are set to be $\rho=0.05$, $0.1$, $1.5\rho_{0}$ and $2\rho_{0}$ for (a), (b), (c) and (d), respectively. At
the density $\rho=0.05$ [in Fig.7.(a)], the two curves of
$\delta\emph{F}$ for the ADG and angle-averaged gap states are
very close to each other, indicating the effect of angle
dependence of the pairing gap is quite small at low densities. When the density increases, the difference of $\delta\emph{F}$ for
the ADG and angle-averaged gap states increases rapidly, implying that the angle dependence of the pairing gap is more
important at higher densities. As the Fermi energy
$\emph{E}_{F}\propto\rho^{\frac{2}{3}}$, the
value of $\frac{\Delta}{\emph{E}_{F}}$ is thus small at high densities. In this case, the
summations over ${k}^{'}$ in the gap equation (9) concentrate near
the average Fermi surface (i.e., the contribution to superfluidity
from the Cooper pairs around the average Fermi surface is
dominant). A little separation of the neutron and proton Fermi surfaces
$\delta\mu$ may suppress the superfluidity strongly. In the ADG
configuration, the angle dependence can reduce the suppression.
However, at low densities, the value of $\frac{\Delta}{\emph{E}_{F}}$
gets large. Thus the contribution to superfluidity from the Cooper
pairs near the average Fermi surface is no longer as important as
that at high densities. Since the angle dependence mainly
increases the pairing probability around the average Fermi
surface, the effect of the angle dependence becomes weak at low
densities.

\section{Summary and Outlook}
The fermionic condensation in asymmetric nuclear matter leads to
superconducting states which spontaneously break the spatial
symmetries (such as FFLO and DFS states). The quasiparticle
spectrum behaves as an isotropic one and the angle dependence of
the pairing gap should be reconsidered. In this work we propose an
axi-symmetric angle dependent gap state in which the isotropic
symmetry is broken in isospin-asymmetric nuclear matter, and
compare with the angle-averaged gap state. It is shown the ADG state is more
favored than the angle-averaged gap state for large asymmetry at low temperature, and the
differences of both the gap values and the free energies between
the two kinds of states get small with increasing temperature. At
temperature $\beta^{-1}=0.5$ MeV with density $\rho_{0}$, the
maximal differences of $\Delta_{0}(k_{F})$ and $\delta\emph{F}$ between the ADG state and angle-averaged gap state are about 22
and 35 percent, respectively. The differences get larger at
higher densities for $\beta^{-1}=0.5$ MeV. In the ADG state, the
neutron and proton deformed Fermi spheres increase the pairing
probability along the axis of symmetry breaking near their average
Fermi surface. The effect of the coupling among different $m_{j}$
gap components is also investigated in this work and we find
the coupling dominates the main contribution to the mechanism of the
ADG state.

The ADG state vanishes at the critical value $\alpha_{c}$, where
the angle-averaged gap vanishes. And the phase transition from the
ADG state to the normal state is of the second order. When
temperature goes up, $\alpha_{c}$ rises and the effect of
angle dependence of pairing gap becomes weak. In a certain region
of $\alpha$ the latent heat has an anomalous negative sign, which
is consistent with the result if Ref.\cite{Sdd2}. However, this does
not affect the stability of the ADG state, since its energy budget
is dominated by the pair-condensation energy.

In the ADG state, the symmetry is broken spontaneously. It is
different from that in the FFLO state, where the symmetry is broken by the
collective motion of the cooper pairs (the translation and
rotational symmetries are both broken). The translation symmetry is
maintained in the ADG state. The deformation of the neutron/proton Fermi sphere
in the ADG state is similar to the DFS configuration, however, the
mechanisms are different. In the DFS state the symmetry breaking
corresponds to the deformed Fermi surface, while in the ADG state
the symmetry breaking results from the angle dependence of the pairing gap.
As is well known, the continuous symmetry breaking leads to collective
excitations with vanishing minimal frequency (Goldstone's
theorem). The breaking of rotational symmetry, which corresponds to
the anisotropic $D^{2}(\textbf{k})$ in the ADG state, may imply new
collective bosonic modes in asymmetric nuclear matter. However,
the true ground state could be a combination of the ADG state and
the FFLO state, we should consider the ADG state with the cooper
pair momentum together which is in progress.

\section*{Acknowledgments}
{The work is supported by the 973 Program of China under No.
2013CB834405, the National Natural Science Foundation of China
(No. 11175219), and the Knowledge Innovation Project(No.
KJCX2-EW-N01) of the Chinese Academy of Sciences.}
\appendix
\section*{Appendix}
We present here the main steps of the elimination of the second
term in Eq.(36) by using the gap equation (9). The elements of the
density matrix of the particles in condensate are,
\begin{eqnarray}
\nu_{\sigma_{1},\sigma_{2}}(\textbf{k})=\frac{\Delta_{\sigma_{1},\sigma_{2}}(\textbf{k})}{2\sqrt{\xi_{\textbf{k}}^{2}+D^{2}(\textbf{k})}}[1-f(E_{\textbf{k}}^{+})-f(E_{\textbf{k}}^{-})].
\end{eqnarray}
The second term of Eq.(36) is written as
\begin{eqnarray}
&&\sum_{\textbf{k},\textbf{k}^{'}}\sum_{\sigma_{1},\sigma_{2},\sigma_{1}^{'},\sigma_{2}^{'}}
<\textbf{k}\sigma_{1},-\textbf{k}\sigma_{2}\mid V\mid\textbf{k}^{'}\sigma_{1}^{'},-\textbf{k}^{'}\sigma_{2}^{'}>
\nu^{\dag}_{\sigma_{2},\sigma_{1}}(\textbf{k})
\nu_{\sigma_{1}^{'},\sigma_{2}^{'}}(\textbf{k}^{'})\nonumber\\&&=
\sum_{\textbf{k},\textbf{k}^{'}}\sum_{\sigma_{1},\sigma_{2},\sigma_{1}^{'},\sigma_{2}^{'}}
<\textbf{k}\sigma_{1},-\textbf{k}\sigma_{2}\mid V\mid\textbf{k}^{'}\sigma_{1}^{'},-\textbf{k}^{'}\sigma_{2}^{'}>
\frac{\Delta^{\dag}_{\sigma_{2},\sigma_{1}}(\textbf{k})}{2\sqrt{\xi_{\textbf{k}}^{2}+D^{2}(\textbf{k})}}[1-f(E_{\textbf{k}}^{+})-f(E_{\textbf{k}}^{-})]\nonumber\\
&&\times
\frac{\Delta_{\sigma_{1}^{'},\sigma_{2}^{'}}(\textbf{k}^{'})}{2\sqrt{\xi_{\textbf{k}^{'}}^{2}+D^{2}(\textbf{k}^{'})}}[1-f(E_{\textbf{k}^{'}}^{+})-f(E_{\textbf{k}^{'}}^{-})]
\nonumber\\&&=
\sum_{\textbf{k},\sigma_{1},\sigma_{2}}
\frac{\Delta^{\dag}_{\sigma_{2},\sigma_{1}}(\textbf{k})}{2\sqrt{\xi_{\textbf{k}}^{2}+D^{2}(\textbf{k})}}[1-f(E_{\textbf{k}}^{+})-f(E_{\textbf{k}}^{-})]\nonumber\\
&&\times
\sum_{\textbf{k}^{'},\sigma_{1}^{'},\sigma_{2}^{'}}<\textbf{k}\sigma_{1},-\textbf{k}\sigma_{2}\mid V\mid\textbf{k}^{'}\sigma_{1}^{'},-\textbf{k}^{'}\sigma_{2}^{'}>
\frac{\Delta_{\sigma_{1}^{'},\sigma_{2}^{'}}(\textbf{k}^{'})}{2\sqrt{\xi_{\textbf{k}^{'}}^{2}+D^{2}(\textbf{k}^{'})}}
[1-f(E_{\textbf{k}^{'}}^{+})-f(E_{\textbf{k}^{'}}^{-})].
\end{eqnarray}
Noting that the second summation over
${k}^{'},\sigma_{1}^{'},\sigma_{2}^{'}$ is
$-\Delta_{\sigma_{2},\sigma_{1}}$ (using the gap equation (9)),
thus
\begin{eqnarray}
&&\sum_{\textbf{k},\textbf{k}^{'}}\sum_{\sigma_{1},\sigma_{2},\sigma_{1}^{'},\sigma_{2}^{'}}
<\textbf{k}\sigma_{1},-\textbf{k}\sigma_{2}\mid V\mid\textbf{k}^{'}\sigma_{1}^{'},-\textbf{k}^{'}\sigma_{2}^{'}>
\nu^{\dag}_{\sigma_{2},\sigma_{1}}(\textbf{k})
\nu_{\sigma_{1}^{'},\sigma_{2}^{'}}(\textbf{k}^{'})\nonumber\\&&=
-\sum_{\textbf{k},\sigma_{1},\sigma_{2}}
\frac{\Delta^{\dag}_{\sigma_{2},\sigma_{1}}(\textbf{k})\Delta_{\sigma_{1},\sigma_{2}}(\textbf{k})}{2\sqrt{\xi_{\textbf{k}}^{2}+D^{2}(\textbf{k})}}
[1-f(E_{\textbf{k}}^{+})-f(E_{\textbf{k}}^{-})]\nonumber\\
&&=-\sum_{\textbf{k}}\frac{Tr[\Delta(\textbf{k})\Delta^{\dag}(\textbf{k})]}{2\sqrt{\xi_{\textbf{k}}^{2}+D^{2}(\textbf{k})}}[1-f(E_{\textbf{k}}^{+})-f(E_{\textbf{k}}^{-})].
\end{eqnarray}
Using the ``unitary" property Eq.(4),
\begin{eqnarray}
\emph{U}&&=\sum_{\sigma\textbf{k}}[\varepsilon_{\textbf{k}}^{(n)}n_{\sigma}^{(n)}(\textbf{k})+\varepsilon_{\textbf{k}}^{(p)}n_{\sigma}^{(p)}(\textbf{k})]\nonumber\\
&&+\sum_{\textbf{k},\textbf{k}^{'}}\sum_{\sigma_{1},\sigma_{2},\sigma_{1}^{'},\sigma_{2}^{'}}
<\textbf{k}\sigma_{1},-\textbf{k}\sigma_{2}\mid V\mid\textbf{k}^{'}\sigma_{1}^{'},-\textbf{k}^{'}\sigma_{2}^{'}>
\nu^{\dag}_{\sigma_{2},\sigma_{1}}(\textbf{k})
\nu_{\sigma_{1}^{'},\sigma_{2}^{'}}(\textbf{k}^{'})
\nonumber\\&&=\sum_{\sigma\textbf{k}}[\varepsilon_{\textbf{k}}^{(n)}n_{\sigma}^{(n)}(\textbf{k})+\varepsilon_{\textbf{k}}^{(p)}n_{\sigma}^{(p)}(\textbf{k})]
-\sum_{\textbf{k}}\frac{2D^{2}(\textbf{k})}{2\sqrt{\xi_{\textbf{k}}^{2}+D^{2}(\textbf{k})}}[1-f(E_{\textbf{k}}^{+})-f(E_{\textbf{k}}^{-})]
\nonumber\\&&=\sum_{\sigma\textbf{k}}[\varepsilon_{\textbf{k}}^{(n)}n_{\sigma}^{(n)}(\textbf{k})+\varepsilon_{\textbf{k}}^{(p)}n_{\sigma}^{(p)}(\textbf{k})]
-\sum_{\textbf{k}}\frac{D^{2}(\textbf{k})}{\sqrt{\xi_{\textbf{k}}^{2}+D^{2}(\textbf{k})}}[1-f(E_{\textbf{k}}^{+})-f(E_{\textbf{k}}^{-})].
\end{eqnarray}


\begin{thebibliography}{90}

\vspace{3mm}

\bibitem{Fn}
A.L.Goodman, Phys. Rev. C {\bf 60}, 014311 (1999), and references therein.
\bibitem{Fn2}
G.R\"{o}pke, A.Schnell, P.Schuck, and U.Lombardo, Phys. Rev. C {\bf 61}, 024306 (2000).
\bibitem{Snm1}
Th.Alm, B.L.Friman, G.R\"{o}pke, and H.Schulz, Nucl. Phys. A {\bf 551}, 45 (1993).
\bibitem{Snm2}
M.Baldo, U.Lombardo, P.Schuck, Phys. Rev. C {\bf 52}, 975 (1995).
\bibitem{Snm3}
E.Garrido, P.Sarriguren, E.Moya de Guerra, and P.Schuck, Phys. Rev. C {\bf 60}, 064312 (1999).
\bibitem{Sdd}
A.Sedrakian, Th.Alm, U.Lombardo, Phys. Rev. C {\bf 55}, R582 (1997).
\bibitem{Sdd2}
A.Sedrakian, U.Lombardo, Phys. Rev. Lett. {\bf 84}, 602 (2000).
\bibitem{Sdd3}
U.Lombardo, P.Nozieres, P.Schuck, H.J.Schulze, A.Sedrakian, Phys. Rev. C. {\bf 64}, 064314 (2001).
\bibitem{Sdd4}
{\O}.Elgaroy, L.Engvik, M.Hjorth-Jensen, and E.Osnes, Phys. Rev. C. {\bf 57}, R1069 (1998).
\bibitem{Sdd5}
A.I.Akhiezer, A.A.Isayev, S.V.Peletminsky, and A.A.Yatsenko, Phys. Rev. C. {\bf 63}, 021304 (2001).
\bibitem{D1}
A.Sedrakian, G.R\"{o}pke, T.Alm, Nucl. Phys. A {\bf 594}, 355 (1995).
\bibitem{D2}
T.Alm, G.R\"{o}pke, A.Sedrakian, and F.Weber, Nucl. Phys. A {\bf 604}, 491 (1996).
\bibitem{Sh1}
S.Typel, G.R\"{o}pke, T.Kl\"{a}hn, D.Blaschke, and H.H.Wolter, Phys. Rev. C. {\bf 81}, 015803 (2010).
\bibitem{Sh2}
S.Heckel, P.P.Schneider, and A.Sedrakian, Phys. Rev. C. {\bf 80}, 015805 (2009).
\bibitem{Sh3}
M.Stein, X.-G.Huang, A.Sedrakian, and J.W.Clark, Phys. Rev. C. {\bf 86}, 062801 (2012).
\bibitem{ff}
P. Fulde and R. A. Ferrell, Phys. Rev. 135 (1964) A550.
\bibitem{lo}
A. I. Larkin and Yu. N. Ovchinnikov, Zh. Eksp. Teor. Fiz. 47
(1964) 1136 [translation, Sov. Phys. JETP 20 (1965) 762].
\bibitem{dfs}
H.M\"{u}ther, and A.Sedrakian, Phys. Rev. Lett. {\bf 88}, 252503 (2002).
\bibitem{ffn}
A.Sedrakian, Phys. Rev. C. {\bf 63}, 025801 (2001).
\bibitem{dfsn}
H.M\"{u}ther, and A.Sedrakian, Phys. Rev. C. {\bf 67}, 015802 (2003).
\bibitem{aap}
M.Baldo, I.Bombaci, and U.Lombardo, Phys. Lett. B. {\bf 283}, 8 (1992).
\bibitem{3pf2}
M.Baldo, J.Cugnon, A.Lejeune, and U.Lombardo, Nucl. Phys. A {\bf 536} 349 (1992).
\bibitem{bhf1}
U.Lombardo, H.-J.Schulze, and W. Zuo, Phys. Rev. C. {\bf 59}, 2927 (1999).
\bibitem{bhf2}
M.Baldo, U.Lombardo, H.-J.Schulze, and Zuo Wei, Phys. Rev. C. {\bf
66}, 054304 (2002).
\bibitem{scr1}
Caiwan Shen, U.Lombardo, P.Schuck, Phys. Rev. C. {\bf 71}, 054301 (2005).
\bibitem{scr2}
L. G. Cao, U.Lombardo, P.Schuck, Phys. Rev. C. {\bf 74}, 064301
(2006).
\end{thebibliography}
\end{document}